\documentclass[conference]{IEEEtran}
\IEEEoverridecommandlockouts
\usepackage{hyperref}
\usepackage{cite}
\usepackage{todonotes}
\usepackage{amsmath,amssymb,amsfonts}
\usepackage{algorithmic}
\usepackage{graphicx}
\usepackage{textcomp}
\usepackage{xcolor}

\usepackage{subcaption}
\usepackage{booktabs}

\def\BibTeX{{\rm B\kern-.05em{\sc i\kern-.025em b}\kern-.08em
    T\kern-.1667em\lower.7ex\hbox{E}\kern-.125emX}}
\begin{document}

\title{Energy Consumption and Performance of \\Heapsort in Hardware and Software*\\
\thanks{This work is supported by the Innovation Fund Denmark for the project DIREC (9142-00001B) and by CERCIRAS Cost Action CA19135.}
}

\author{\IEEEauthorblockN{Maja H.\ Kirkeby, Thomas Krabben, Mathias Larsen,\\ Maria B.\ Mikkelsen, Mads Rosendahl, Martin Sundman}
\IEEEauthorblockA{\textit{Department of People and Technology} \\
\textit{Roskilde University}\\
Roskilde, Denmark \\
majaht@ruc.dk, krabben@ruc.dk, mamaar@ruc.dk,\\ mariabm@ruc.dk, madsr@ruc.dk, sundman@ruc.dk}
\and
\IEEEauthorblockN{Tjark Petersen, Martin Schoeberl}
\IEEEauthorblockA{\textit{DTU Compute} \\
\textit{Technical University of Denmark}\\
Lyngby, Denmark \\
s186083@student.dtu.dk, masca@dtu.dk}
}

\maketitle

\begin{abstract}

In this poster abstract we will report on a case study on implementing the Heapsort algorithm in hardware and software and comparing their time and energy consumption. Our experiment shows that the Hardware implementation is more energy efficient, but slower than the Software implementation due to a low clock frequency. It also indicate that the optimal degree of parallelization differs when optimizing for time compared to optimizing for time.
\end{abstract}


\vspace{-0.1cm} 
\section{Background}
Programs running on a general purpose computer consume a considerable amount
of energy. Some programs can be translated into hardware and executed on an field-programmable gate array (FPGA).
The study presented here is a first step in studying  whether and how it is possible to reduce energy consumption of IT systems by moving algorithms from software into hardware (FPGAs). 
In this first study, the focus is on implementations of a sorting algorithm: In later works the framework will be extended to path finding algorithms

\vspace{-0.1cm} 
\section{Aims/purpose}
We will do this by investigating the classic sorting algorithm Heapsort, e.g., see~\cite[Chp.6]{Cormen2009}, implementing it in software and hardware and comparing their performance and energy-efficiency. Such results are essential for reducing energy and time consumption in both data centers as well as embedded systems. The choice of Heapsort for this analysis is based on a number of factors.
Heapsort has a known worst and best case complexity of $O(n\log_2(n))$ with limited data dependency on execution time; when generalizing the binary heap to a $k$-heap the complexity is $O(n\log_k(n))$. Sorting can be done in memory without using extra dynamic data. 
The algorithm also allows us to parallelize certain elements of the algorithm and thus facilitating an investigation of the effects
parallel evaluation can have on resource and energy consumption.

\vspace{-0.1cm} 
\section{Methods}
\begin{figure*}[!t]
\centering
\includegraphics[width=0.3\textwidth]{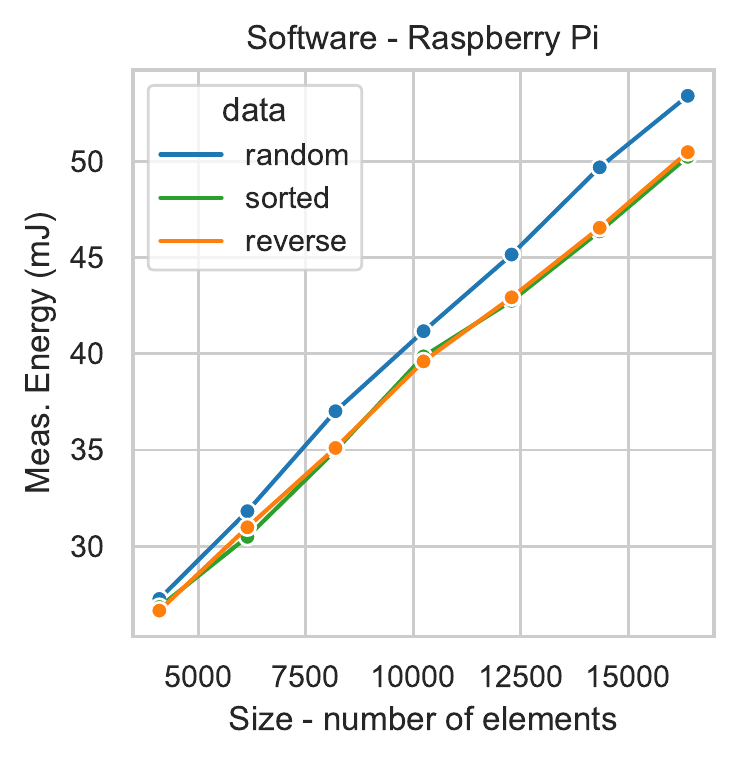}~%
\includegraphics[width=0.3\textwidth]{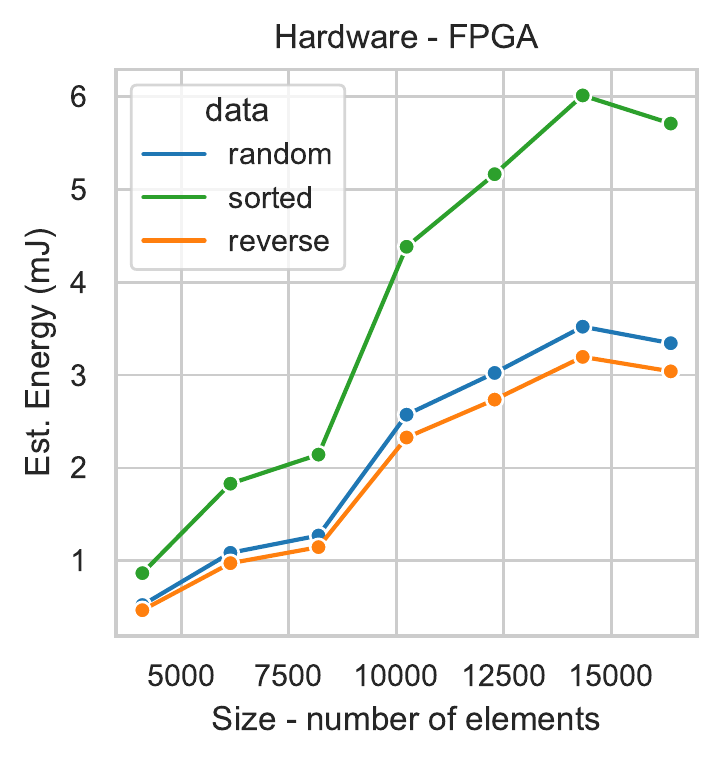}~%
\includegraphics[width=0.3\textwidth]{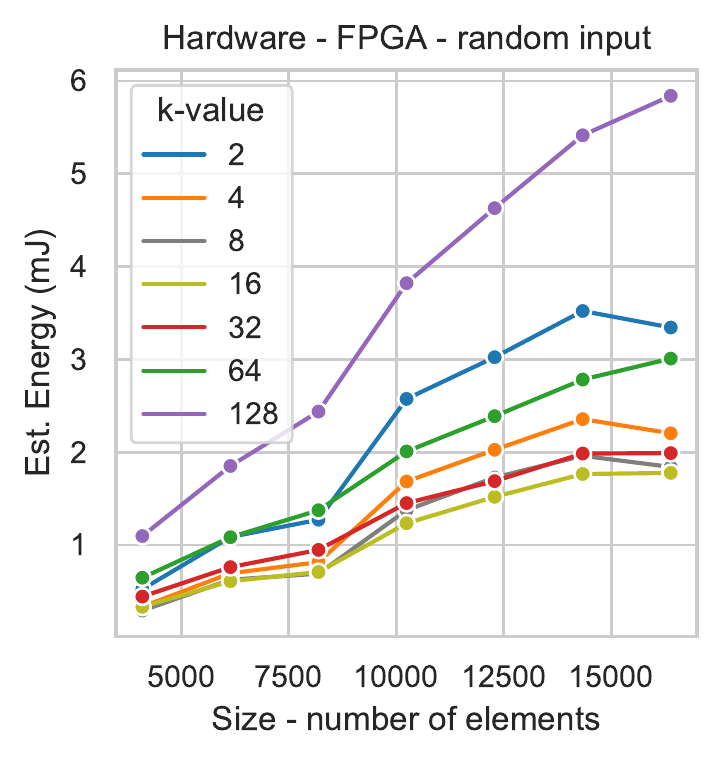}

\caption{Measured and Estimated Energy Consumption (in mJ) for Raspberry Pi and FPGA, respectively. The third subfigure shows Estimated Energy Consumption of sorting the randomly ordered list for different values of $k$. 
}
    \label{fig:results}
    \vspace{-0.3cm} 
\end{figure*}
\paragraph{Software Implementation}
The software implementation of Heapsort is written in C and is based on the standard solution from Rosetta Code\footnote{\url{http://www.rosettacode.org/wiki/Sorting_algorithms/Heapsort}} which follows the description in~\cite[Chp.6]{Cormen2009}. The solution implements a 
max-heap. For easy comparison with the hardware implementation, the solution is altered so that the input list is a variable in the compiled program and when the program is executed the list elements are inserted into a heap structure. To ensure the max-heap property when inserting or removing a new element, all the nodes/elements violating the heap order are swapped while traversing the heap until each node is larger than its children.
\paragraph{Hardware Implementation}

The hardware implementation consists of a heap module and a state machine. The latter first inserts all input elements from an on-chip memory into the heap and afterwards writes them in sorted order from the heap back into the on-chip memory. The heap module has a variable tree order $k$. It maintains the max-heap property the same way the software implementation does.

In the hardware solution, we can exploit parallelism in two cases: when considering the memory layout, and when finding the maximum between a node and its children. In both cases, more parallelism can be exploited by increasing the tree order $k$. The heap module uses its own memory to store values while they reside in the heap. All sibling nodes are spread over $k$ memory banks. This gives us single cycle access to all children of a given node at the same time. Finding the maximum between a node and its children can be parallelized using a reduction tree. Like in a tournament, we find the maximum between pairs, halving the candidates in each cycle until only one winner is left. This reduces the number of cycles from $k$ to $\log_2(k)$ for this operation.

\paragraph{Measuring and Estimating Time, Power, and Energy}
Because the runtime depends on (i) the number of elements of Heapsort's input lists, we have varied the number of elements in the input lists, i.e., the size, from 4096 up to 16384 elements; and (ii) the order of these elements, we have (for each size) generated three different versions of each input list: one with random order, one with sorted elements and one with reverse sorted elements. We have used the same lists as input to both the software and hardware implementations.

The software Heapsort implementation in C is executed on a standard \emph{Raspberry Pi 4 computer Model B 4GB RAM and a 1.5 GHz 64-bit quad-core ARM Cortex-A72 processor} using the \verb+gcc+ compiler\footnote{Version: gcc (Raspbian 8.3.0-6+rpi1) 8.3.0} with flag \verb+-O2+. We have created one program per input list.
We measure the energy consumption for the entire Raspberry Pi using a programmable power supply \emph{Siglent SPD3303X-E Linear DC 3CH}. To obtain average power dissipation and average execution times, we 
executed
each program 2000 times with 20 seconds sleep in between. This approach 
allowed 
us to measure the execution time 2000 times for each program and at the same time sampling the power dissipation externally. 

The execution time of the hardware Heapsort implementation is deterministic for a given data set. 
We can therefore identify the exact number of cycles  for each test case by running a simulation. 
For this purpose, we have used the Chiseltest framework~\cite{chiseltest}. 
We have synthesized the design to the Digilent Basys 3 Artix-7 FPGA board for all configurations.
The number of clock cycles together with the 100MHz clock frequency of the Basys 3 board give us exact run time results, which are provided in Table~\ref{tab:results}. 
To obtain energy estimates, we have used Xilinx Vivado's built-in power reporting tool; 
this provides rough power dissipation estimates for the hardware implementations. 
%
\vspace{-0.3cm} 
\section{Results}
The execution time increases steadily relative to the number of inputs for both implementations, see Table~\ref{tab:results} with the results for a standard binary heap, i.e., $k=2$.
The hardware implementation runs slightly faster for input sizes $4096$ and $6144$, whereas the software implementation runs faster for larger input sizes, see ``Time" column in Table~\ref{tab:results}. This is expected since Raspberry processor runs 15 times faster, when they both execute the same operations (sequentially).

Figure~\ref{fig:results} show that the energy consumption of the implementations increase with the number of elements to be sorted. For the software implementation, the measured power dissipation had only minor variations and the execution is an accurate proxy of the energy consumption~\ref{tab:results}; this aligns with the findings for sequential Haskell programs~\cite{Lima2016}.
While the execution time for the hardware implementations is predictable, its estimated energy consumption is less predictable; this could be due to the uncertainty introduced by the rough estimates of the power dissipation.  
%


When considering $k$-heaps, our experiments showed an increase of performance of the hardware implementation when $k$ increased. This aligns with the complexity of the algorithm and especially for hardware this is expected, since a higher $k$ facilitates a higher degree of parallelism. 
While $k=128$ is the better choice when optimizing the hardware implementation for time, Figure~\ref{fig:results} demonstrates that $k=16$ is optimal when optimizing for energy (assuming at least relative precision of the power dissipation estimates).


\begin{table}[b]
  \centering
  \vspace{-0.2cm} 
  \caption{The avg. execution time and energy consumption of hardware with $k=2$ (FPGA) over software (Raspberry Pi) executions for random input, together with the FPGA's (non)-improvements over the Raspberry Pi.}
\begin{tabular}{crrrrrr}
    \toprule
     &\multicolumn{2}{c}{Time (ms)} & \multicolumn{2}{c}{Energy Cons.~(mJ)} &  \multicolumn{2}{c}{FPGA Improvement}\\
Size&Rasp.~Pi&FPGA&Rasp.~Pi&FPGA& Time & Energy \\
\midrule
4096 	 & 8.001 	 & 5.386 	 & 27.27 	 & 0.522 	 & 1.486 	 & 52.241 \\
6144 	 & 9.263 	 & 8.479 	 & 31.824 	 & 1.085 	 & 1.092 	 & 29.331 \\
8192 	 & 10.696 	 & 11.665 	 & 37.015 	 & 1.271 	 & 0.917 	 & 29.123\\
10240 	 & 12.023 	 & 14.963 	 & 41.17 	 & 2.574 	 & 0.804 	 & 15.995\\
12288 	 & 13.137 	 & 18.322 	 & 45.148 	 & 3.023 	 & 0.717 	 & 14.935\\
14336 	 & 14.604 	 & 21.737 	 & 49.678 	 & 3.521 	 & 0.672 	 & 14.109\\
16384 	 & 15.749 	 & 25.138 	 & 53.39 	 & 3.343 	 & 0.627 	 & 15.971\\
\bottomrule
\end{tabular}
    \label{tab:results}
\end{table}


\vspace{-0.1cm} 
\section{Conclusion}
In this poster abstract, we have compared performance and energy efficiency of software and hardware implementations of Heapsort. We find that while the hardware implementation runs slower on an FPGA board compared to running the software implementation on a Rasperry Pi; the hardware implementation is more energy efficient. We also note that the optimal degree of parallelization of the hardware implementation differs when optimizing for energy versus optimizing for time.


\bibliography{bibliography.bib}
\bibliographystyle{plain}

\end{document}